\documentclass[fleqn,usenatbib]{mnras}

\usepackage{newtxtext,newtxmath}

\usepackage[T1]{fontenc}

\DeclareRobustCommand{\VAN}[3]{#2}
\let\VANthebibliography\thebibliography
\def\thebibliography{\DeclareRobustCommand{\VAN}[3]{##3}\VANthebibliography}

\usepackage{tikz}
\usepackage{amsmath}
\usepackage{hyperref}
\usepackage{cleveref}
\usepackage{graphicx}
\usepackage{tikz-3dplot}
\usepackage[acronym]{glossaries}
\usepackage{enumitem}

\usepackage{physics}
\usepackage{siunitx}

\usepackage{nicefrac} 
\usepackage{float} 

\newcommand{\Rsun}{\textrm{R}_\mathrm{\odot}}
\newcommand{\Msun}{\textrm{M}_\mathrm{\odot}}
\newcommand{\Mearth}{\textrm{M}_\mathrm{\oplus}}
\newcommand{\Mj}{\textrm{M}_\mathrm{J}}

\newcommand{\PreserveBackslash}[1]{\let\temp=\\#1\let\\=\temp}

\renewcommand{\qc}{\,\text{,}}
\newcommand{\qs}{\,\text{.}}

\tdplotsetmaincoords{60}{110}
\pgfmathsetmacro{\rvec}{1.0}
\pgfmathsetmacro{\thetavec}{30}
\pgfmathsetmacro{\phivec}{60}

\newacronym{lisa}{LISA}{Laser Interferometer Space Antenna}
\newacronym{gw}{GW}{gravitational wave}
\newacronym{gpu}{GPU}{graphics processing unit}
\newacronym{esa}{ESA}{European Space Agency}
\newacronym{tdi}{TDI}{time-delay interferometry}
\newacronym{fir}{FIR}{finite impulse response}
\newacronym{iir}{IIR}{infinite impulse response}
\newacronym{psd}{PSD}{power spectral density}
\newacronym{asd}{ASD}{amplitude spectral density}
\newacronym{ldc}{LDC}{LISA Data Challenge}
\newacronym{ssb}{SSB}{Solar system's barycenter}
\newacronym[firstplural=moveable optical sub-assemblies]{mosa}{MOSA}{moveable optical sub-assembly}

\title{Bayesian Characterisation of Circumbinary Exoplanets with {\it LISA}}

\author[M. L. Katz et al.]{
Michael L. Katz,$^{1}$\thanks{E-mail: michael.katz@aei.mpg.de}
Camilla Danielski,$^{2,3}$
Nikolaos Karnesis,$^{4}$ 
Valeriya Korol,$^{5}$
Nicola Tamanini,$^{6}$
\newauthor
Neil J. Cornish,$^{7}$
Tyson B. Littenberg,$^{8}$
\\
$^{1}$Max-Planck-Institut f\"ur Gravitationsphysik, Albert-Einstein-Institut,  Am M\"uhlenberg 1, 14476 Potsdam-Golm, Germany\\
$^{2}$Instituto de Astrofísica de Andalucía, CSIC, Glorieta de la Astronomía, 18008, Granada, Spain\\
$^{3}$Sorbonne Universit\'es, UPMC Universit\'e Paris 6 et CNRS, UMR 7095, Institut d'Astrophysique de Paris, 98 bis bd Arago,
75014 Paris, France\\
$^{4}$ Department of Physics, Aristotle University of Thessaloniki, Thessaloniki 54124, Greece \\
$^{5}$Institute for Gravitational Wave Astronomy \& School of Physics and Astronomy, University of Birmingham, Birmingham, B15 2TT, UK\\
$^{6}$Laboratoire des 2 Infinis - Toulouse (L2IT-IN2P3), CNRS, UPS, F-31062 Toulouse Cedex 9, France\\
$^{7}$eXtreme Gravity Institute, Department of Physics,
Montana State University, Bozeman, Montana 59717, USA \\
$^{8}$NASA Marshall Space Flight Center, Huntsville, Alabama 35811, USA
}

\date{Accepted XXX. Received YYY; in original form ZZZ}

\pubyear{2022}

\begin{document}
\label{firstpage}
\pagerange{\pageref{firstpage}--\pageref{lastpage}}
\maketitle

\begin{abstract}
The Laser Interferometer Space Antenna ({\it LISA}) will detect and characterize $\sim10^4$ Galactic Binaries consisting predominantly of two White Dwarfs (WD). An interesting prospect within this population is a third object--another WD star, a Circumbinary Exoplanet (CBP), or a Brown Dwarf (BD)--in orbit about the inner WD pair. We present the first fully Bayesian detection and posterior analysis of sub-stellar objects with {\it LISA}, focusing on the characterization of CBPs. We used an optimistic astrophysically motivated catalogue of these CBP third-body sources, including their orbital eccentricity around the inner binary for the first time. We examined Bayesian Evidence computations for detectability, as well as the effects on the posterior distributions for both the inner binary parameters and the third-body parameters. We find that the posterior behavior bifurcates based on whether the third-body period is above or below half the observation time. Additionally, we find that undetectable third-body sources can bias the inner binary parameters whether or not the correct template is used. We used the information retrieved from the study of the CBP population to make an initial conservative prediction for the number of detectable BD systems in the original catalogue. We end with commentary on the predicted qualitative effects on {\it LISA} global fitting and Galactic Binary population analysis. The procedure used in this work is generic and can be directly applied to other astrophysical effects expected within the Galactic Binary population.
\end{abstract}





\section{Introduction}
The Laser Interferometer Space Antenna ({\it LISA}) \citep{LISAMissionProposal}, a future space-based gravitational wave detector set to launch in the mid-2030s, will add new gravitational-wave observations to the high-frequency observations of ground-based gravitational-wave detectors \citep[e.g.,][]{LIGOScientific:2018mvr, LVK2018LivingReview}. {\it LISA} will observe in the milliHertz regime of the spectrum, an area where many different types of astrophysical sources are expected to be present \citep{LISAastroWP}. An important class of {\it LISA} sources are compact Galactic Binaries (GB): a binary consisting of two compact objects with short period orbits located within the Milky Way or surrounding close satellites.
Typically, GBs will include two White Dwarfs\footnote{In this paper, we refer to double White Dwarf binaries as both `GB' and `DWD.'}, but some GBs may contain one or more stellar-origin black holes or neutron stars \citep{LISAMissionProposal}. There are roughly $\sim10^7$ GBs in the Milky Way emitting GW in the {\it LISA} frequency band \citep[e.g.,][]{Korol:2021pun}. Of these, $\sim10^4$ will be individually resolvable with signal-to-noise ratios (S/N) of $\sim10-10^3$ \citep[e.g.,][]{Timpano:2005gm,Korol2017,Cornish:2017vip,Breivik2020,Lamberts2019}. Electromagnetic observations of double White Dwarf binaries have found $\gtrsim20$ \citep{Kupfer:2018jee} individually resolvable GB {\it LISA} sources based on their electromagnetically determined properties. These systems are the so-called `Verification Galactic Binaries'. The remaining unresolved GBs at low S/N will combine into a confusion background that is expected to lie above the low-frequency part of the {\it LISA} sensitivity curve \citep[e.g.,][]{Robson:2017ayy}.

GBs are quasimonochromatic sources that evolve slowly over the duration of the {\it LISA} observation. These systems will be long-lived and present in {\it LISA} data from the beginning to the end of its observation.
They emit gravitational waves deep in the inspiralling stage, from tens of thousand to million years prior to merger.
The overlap of resolvable and unresolvable GBs, in addition to the rest of the other {\it LISA} sources, will greatly complicate the analysis of the LISA data. Proper modelling of the {\it LISA} spacecraft orbits and the observed {\it LISA} instrumental noise will also be necessary for accurate analysis, even in the presence of expected non-stationary and non-Gaussian noise effects. The main method suggested to solve this problem is using global fitting techniques that fit the parameters of many astrophysical sources and noise properties simultaneously~\citep{Cornish:2005qw}. The global fit of GBs specifically has been examined recently in \citet{Littenberg:2020bxy}. This analysis was performed assuming a basic set of astrophysical GBs. It proved the capabilities of global fit analyses and discussed further development of the method to include more realistic effects of the GB population and {\it LISA} detector setup. 

Many science questions can be addressed with {\it LISA} observations of GBs on different scales ranging from sub-stellar/stellar to the entire Galaxy \citep{LISAastroWP}. For example, at the population level, the characteristics of the observed confusion background from GBs and the spatial distribution of resolved ones can help to understand the stellar distribution and properties of the Milky Way Galaxy \citep[][]{Benacquista2006,Adams2012,Breivik:2019oar,Korol2019,Wilhelm2021,Georgousi2022}. The population of resolvable GBs can also be used to constrain tests of General Relativity (GR) \citep{Littenberg:2018xxx}.  Resolvable source observations, both for individual sources and the population as a whole, can shed light on the complex astrophysics of multiple-object systems \citep[e.g.,][]{Kremer:2017xrg}.

Roughly $13\%$ of low-mass stellar systems contain three or more stars \citep{Tokovinin2014, Fuhrmann2017}. Low-mass binaries with periods of less than three days are expected to exist in hierarchical systems at a fraction of $\sim96\%$ \citep{Tokovinin:2006jm, Qian:2011yp}. The predicted detection rate of these systems by {\it LISA} is uncertain; however, it is reasonable to expect these systems to play a large role in the observed GB population. Astrophysical effects, such as Kozai-Lidov oscillations, can cause a hardening of the inner binary, driving it to closer separations \citep{Kozai1962,Lidov1962}. This may enlarge the fraction of observed GB systems containing a third-body \citep{Thompson2011, Fang2018, Toonen:2017yct, Fabrycky:2007xd}. 
Within the sub-stellar objects regime, analytical studies by \citet{MartinTriaud2016} showed that the Kozai-Lidov evolution rarely affects the inner binary in systems hosting a tertiary planet due to its low mass. However, in the brown-dwarf regime there is a significant change in the Kozai-Lidov behaviour for tight triple systems, where the ratio between the outer and inner binary semi-major axes is less than 15 \citep{MartinTriaud2016}. 

In this paper, we are specifically interested in the observations of hierarchical triple systems where a Circumbinary Exoplanet (CBP) is in orbit around an inner binary consisting of two WDs. We examine the posterior distributions stemming from a population of these sources (see Section~\ref{sec:pop}). We also examine observations of the population of circumbinary Brown Dwarfs (BD).  As we will discuss, the number of potentially detectable\footnote{By `detectable' we mean `distinguishable' according to the Bayesian Evidence ratio (see Section~\ref{sec:bayes}).} BDs in the population is much higher than CBPs, making a full analysis of the BD systems computationally costly. Therefore, we focus our detailed study on CBP systems and provide general comments on the BD systems in the catalogue which will be examined in more detail in the future (\citeauthor{Danielski-in-prep}). The cut off between the two object classes was chosen to be 16 $\Mj$ (as discussed in \citealt{Spiegel2011}), instead of 13 $\Mj$, the deuterium burning limit.

These populations have been examined previously in \citet{ Robson:2018svj,Tamanini:2018cqb, Danielski:2019rvt, Tamanini:2019usx, Kang:2021bmp}. The statistical Information Matrix\footnote{The `Information Matrix' is also referred to as the `Fisher Information Matrix'. In order to foster a welcoming community in our field, we have not used this term due to its namesake's connection with the science of and personal advocacy for Eugenics \citep[e.g.,][]{Fishercitation}.} was used to analyze the detection and characterization of CBPs and BDs with {\it LISA} in \citet{Tamanini:2018cqb} and \citet{Danielski:2019rvt}. Recently, the Information Matrix was also deployed to study CBP detection and characterization by {\it Taiji}, a space-based interferometer comparable in scope to {\it LISA}~\citep{Kang:2021bmp}. In~\citet{Robson:2018svj}, fast frequency-domain waveforms containing the third-body effect were constructed and an initial Bayesian analysis was performed for systems with three stellar-mass objects in a hierarchical triple. A key finding from this work was initial relations comparing different regimes of $P_2$, the period of the perturber, compared to $T_\text{obs}$, the {\it LISA} observation time. With $P_2/T_\text{obs} \gg 1$, the effect of the perturber is not detectable. Near $P_2/T_\text{obs}\lesssim10$, the companion object can create a detectable doppler shift in the waveform. As the period of the third body decreases so that $P_2/T_\text{obs} \lesssim 1$, the eccentricity and period of the outer orbit can be characterized. While these results relate to systems with a third body that is of similar stellar mass to the inner GB constituents, we will show that similar relations of $P_2$ to $T_\text{obs}$ govern the posterior behavior of CBP systems. Specifically, there is a clear difference between behavior above and below $P_2=T_\text{obs}/2$, which represents the Nyquist criterion for sampling of the full third-body orbit. 

Here, we expand on this previous work by performing a full Bayesian analysis, including posterior and Evidence estimation, on the observability and characterization of individual instances of these triple systems by {\it LISA}. We will examine how many systems, stemming from a coherent and astrophysically motived population, produce a waveform where the effect of the third body is statistically significant compared to the base inner-binary waveform. In addition to purely detecting and characterizing these sources, we strive to illuminate issues that may arise from fitting base two-body GB templates to true source waveforms that include the presence of the third body. The third-body-inclusive waveforms used in this work are also the first to include eccentric third-body orbits in the waveform description when analysing CBP systems in LISA. 

In Section~\ref{sec:pop}, we describe the population of sources used in this work. Sections~\ref{sec:gbwave} and~\ref{sec:bayes} detail the GB waveform model and Bayesian analysis techniques employed, respectively. The results of our analysis pipeline are given in Section~\ref{sec:catalogue_analysis}. In Section~\ref{sec:discussion}, we discuss these results within the current context of GB analysis with {\it LISA}. We conclude our remarks in Section~\ref{sec:conclude}. Throughout this work, we use geometrical units with $G=c=1$. 

\section{Population of circumbinary sub-stellar objects orbiting detached double white dwarf binaries} \label{sec:pop}


In this work we consider hierarchical triple systems consisting of an inner detached double white dwarf (DWD) binary and a tertiary sub-stellar object (SSO).
Specifically, we focus on DWD systems only as they will be the most numerous among the other GBs accessible with {\it LISA} \citep{LISAastroWP} and thus more likely to provide triple systems. 
As a first step we assemble the DWD population, constituting the inner component of the hierarchical triple system. Next, we inject the SSO population into the already formed DWD population i.e., neglecting co-evolution of the stellar binary and the third object. 

\subsection{Inner DWD binaries} \label{sec:dwd_pop}

We follow a binary population synthesis approach to obtain a representative Galactic DWD population. Specifically, we assemble our mock population based on the DWD evolution model constructed using population synthesis code {\sc SeBa} \citep{PZ96,Nelemans01a,Toonen12}, publicly available as part of the {\sc AMUSE} environment \citep{amuse}. 
The choice of the fiducial model is motivated by the fact that it yields the DWD space density in a good agreement with that derived from currently available spectroscopically-selected samples, and it reproduces the general trend of the observed DWD mass ratio distribution, which peaks at around unity \citep{Toonen12,Toonen2017}.  We refer for a detailed description of this model to \citet{Toonen12} and we describe below only its main characteristics. 

The initial (i.e., zero-age main-sequence) population is assembled with a Monte Carlo technique. The initial binary fraction is assumed to be of 50\%, and metallicity is set to the Solar value for all binaries. The mass of the primary star -- initially more massive of the two -- is sampled between 0.95 - 10\,M$_\odot$, according to the initial mass function of \citet{KroupaIMF}. The mass of the secondary star is defined such to obtain a flat mass ratio distribution between 0 and 1 \citep[e.g.,][]{Duchene2013}. Binaries' semi-major axes  are sampled from a log-uniform distribution extending up to $10^6\,$R$_\odot$ \citep{Abt1983,Raghavan2010,Duchene2013}, while orbit eccentricities are sampled from a thermal distribution \citep{Heggie1975}.  

{\sc SeBa} evolves the obtained initial population until both stars become WDs following prescriptions for processes shaping the binary evolution path; these include mass and angular momentum transfer, common envelope evolution, magnetic braking, and gravitational radiation \citep[][and references therein]{PZ96, Toonen12}. It is important to mention that to obtain a close DWD pair emitting GWs in {\it LISA} today, binaries typically experience two common envelope phases \citep[e.g.,][]{Nelemans01a}. Thus, the assumption about the common envelope evolution is of particular importance, and has been found to be the largest source of uncertainty in binary population synthesis models. 
In our fiducial $\gamma\alpha$ model, the first common envelope phase is typically described by the $\gamma$ formalism (based on the angular momentum balance equation that allows for both shrinkage and the widening of the orbit), while the second is described by the $\alpha$ formalism (based on the energy balance equation that always leads to the shrinkage of the orbit). This scenario has been derived based on the reconstruction of the evolutionary paths of individually observed DWD binaries \citep{Nelemans2000,Nelemans2005,van06}.

Next, we distribute DWDs in a Milky Way-like galaxy as in \citet{Korol2019}. We adopt a simplified Galactic potential composed of an exponential stellar disc and a spherical central bulge, which contains the majority of stars in the Milky Way. To model the star formation history of the Galaxy, we use the plane-projected star formation rate from a chemo-spectrophotometric model of \citet{BP99}, and we assume the age of the Galaxy to be of 13.5\,Gyr. The obtained integrated star formation history reproduces well the observed Galactic star formation history inferred from single WDs \citep[][]{Fantin2019}. 

Finally, we randomly draw binary initial phases and polarisation angles from a uniform distribution, while the binary inclination angle $\iota_1$ (the `1' indicates the inner binary orbit) is drawn from a uniform distribution in $\cos \iota_1$.

\subsection{Pre-selection of detectable DWDs}



We obtain a catalogue amounting to $\sim 2.9 \times 10^7$ DWDs emitting in the {\it LISA} frequency band. As the next step,
following the methodology described in \cite{Timpano2006, crowder2007,  Nissanke2012eh, Karnesis2021}, we estimate the unresolved confusion foreground signal based on our DWD population, which then allows us to pre-select detectable DWDs based on an S/N criterion. Assuming a {\it LISA} nominal mission duration of 4\,yr (i.e., $T_\text{obs}=4$\,yr) and an S/N threshold of 7, we obtain a subset of $\sim 2.5 \times 10^4$ individually resolvable DWDs to be used in the subsequent steps of our analysis.

\subsection{Tertiary sub-stellar objects}\label{sec:ssoinjection}

The characteristics of the population of injected SSOs follows the optimistic scenario reported in \cite{Danielski:2019rvt}.
 We adopt this optimistic population to have more systems to study in detail. Consequently, the absolute number of {\it LISA} detections that we report here is likely on the high side, and should not be taken as representative of the numbers that {\it LISA} will realistically be able to deliver.
Furthermore, we note that the occurrence rate of the SSOs was based on the observational WD pollution constraints (i.e., 50\% of the DWD population, \citealt{Koester2014}), and that the optimistic scenario was originally defined to have systems with an outer circular orbit. The detection efficiency of circumbinary objects by {\it LISA} will be presented in an upcoming work through a Bayesian study of multiple population scenarios which includes the third-body eccentricity (\citeauthor{Danielski-in-prep}). Here we focus on the data analysis strategy only.

Following \citet{Danielski:2019rvt}, we use a $\log{}_{10}$ uniform distribution for the semi-major axis $a$:  $\log{\mathcal{U}_a}$ ($a_{\rm min}$ - 200 au), and a uniform distribution on the third-body mass $m_c$: $\mathcal{U}_{m_c}$(1 $\Mearth$ - 0.08 $\Msun$).
Novelties with respect to \cite{Danielski:2019rvt} are both the different lower bound $a_{\rm min}$ for the distribution of the semi-major axis, and the inclusion of the eccentricity as an extra parameter defining the orbit of the third body.
The lower bound $a_{\rm min}$ is now physically motivated and is provided by the stability criterion of \citet{2021MNRAS.507.4507B} which, assuming a total DWD mass of 2\ $\Msun$ separated by 1 $\Rsun$ with zero eccentricity, yields $a_{\rm min,1} = 2.39\, \Rsun$. These values have been chosen as representatives among the highest possible values within our DWD population, to provide a conservative condition for the minimum value of $a$.
This condition holds unless the eccentricity of the SSO is so high that the third body enters the Roche limit of one of the WDs at perihelion, in which case it would be destroyed by tidal gravitational forces.
We thus conservatively set the maximum allowed eccentricity $e_{2,\text{max}}$ (the `2' here indicates the outer orbit) using the Roche limit condition \citep{Roche1849} which, given $a$, can simply be computed as
\begin{equation}
    a (1 - e_{2, \rm max}) = 2.44 \cdot {\rm R}_\oplus \cdot \left(\frac{\rho_{\rm WD}}{\rho_{\rm SSO}}\right)^{1/3}  + \frac{\Rsun}{2} \,.
    \label{eq:Roche_lim_condition}
\end{equation}
Here, we assumed co-planar orbits, a typical WD with radius ${\rm R}_\oplus$, and where the last addend on the right hand side of the equation accounts for the maximum separation of the two WDs from the center of mass of the DWD system (we set this to $\Rsun$/2).
The WD density has been taken to be $\rho_{\rm WD}$ = 10$^{9}$ kg/m$^3$, while the density of the third-body $\rho_{\rm SSO}$ has been estimated from a linear fit of the data reported in Figure~1 of \citet{HatzesRauer2015}, with a flat floor imposed below 1 $\Mj$.
Since the eccentricity cannot be lower than zero, the condition in Eq.~\eqref{eq:Roche_lim_condition} at $e_{2,\rm max}=0$ sets an effective minimum semi-major axis $a$ that all SSOs must respect, namely $a_{\rm min,2}$.
For each circumbinary system the minimum of the semi-major axis distribution $a_{\rm min}$ is thus set as the largest separation value chosen between $a_{\rm min,1}$ and $a_{\rm min,2}$.

We now prescribe the distribution of eccentricities for the SSO population.
Since the SSO population is unprobed, and due to a handful of detections of P-type bodies orbiting main sequence binaries, we used the distribution of the eccentricity of single-star planetary systems. Specifically, we combined various distributions found in the literature, as to cover a wide range of SSO-to-binary separations.
When the distribution was unknown, a uniform distribution was applied. In detail, covering a global range from $a_{\rm min}$ - 200 au, we set: 
\begin{itemize}[leftmargin=*]
\setlength\itemsep{0.1em}
    \item $a_{\rm min} < a \leq$ 5 au: uniform distribution $\mathcal{U}_e (0,e_{2, \rm max})$ for both CBPs and BDs.
    \item 5 au $< a \leq$ 200 au: beta distribution  $f(e_2|\alpha, \beta) = \frac{\Gamma(\alpha + \beta)}{\Gamma(\alpha)\Gamma(\beta)} e_2^{\alpha-1}(1-e_2)^{\beta -1}$,  where $\Gamma$ is the gamma function, and where [$\alpha, \beta$] are [30, 200] and [2.30, 1.65] for CBPs and BDs, respectively. These values were presented in \citet{Bowler2020} where the authors studied the eccentricity distribution of directly imaged planets orbiting main sequence, or younger, binaries in the 5-100 au range. Given that the eccentricity increment in stable post-common-envelope systems is very small for separations below 200 au \citep{Veras2011}, we adopted such distributions for our DWDs systems for the whole range up to 200 au.
\end{itemize}

In both cases the eccentricity values were drawn from the respective distributions with an upper limit $e_{2, \rm max}$ chosen to avoid the tidal disruption of the SSO, as given by Eq.~\eqref{eq:Roche_lim_condition}.

\section{Galactic Binary waveforms}\label{sec:gbwave}

GBs are comparatively simple systems to model in terms of their gravitational waveforms, which can be attributed to their slow orbital evolution. Adding in effects that occur on slower timescales than the GB orbital timescale, like accretion or multi-body systems, creates small changes to the phasing of the base waveform. If this phasing difference is significant enough, the astrophysical effect will be detectable. Here, we will discuss the basics of GB waveforms and the process of adding the effect of a third body. We point the interested reader to \citet{Robson:2018svj} for more detailed information on the initial construction of these third-body-inclusive waveforms. The waveforms used here are identical to those used in that paper. 

The two gravitational-wave polarizations from a GB are given by
\begin{align}
 	h_+(t)& = \frac{2\mathcal{M}}{D_L}\left(\pi f_\text{gw}(t)\right)^{2/3}\left( 1 + \cos^2 \iota_1 \right) \cos{\Psi_\text{gw}} \label{eq:polarizations}\\
	&\qquad\qquad\qquad\qquad\qquad\qquad\qquad\qquad\text{and} \nonumber\\
	h_\times(t)& = -\frac{4\mathcal{M}}{D_L}\left(\pi f_\text{gw}(t)\right)^{2/3}\cos{\iota_1} \sin{\Psi_\text{gw}} \qc \label{eq:polarizations2}
\end{align}
where $\mathcal{M}$ is the chirp mass; $f_\text{gw}$ is the instantaneous gravitational wave frequency; $D_L$ is the luminosity distance; $\iota_1$ is the inclination of the inner binary orbit; and $\Psi_\text{gw}$ is the gravitational wave phase over time. The terms in Eq.~\eqref{eq:polarizations} and Eq.~\eqref{eq:polarizations2} are given at the Solar System Barycenter. The phase is related to the integral over frequency: $\Psi_\text{gw}=\phi_0 + 2\pi \int^t f_\text{gw}(t')dt'$, with $\phi_0$ representing an initial arbitrary phase shift.

The instantaneous gravitational-wave frequency is governed by the astrophysical evolution of the source. The timescale over which the frequency changes for a typical GB is slow: $f/\dot{f}\sim10^{-3}\text{Hz}/10^{-17} \text{Hz s}^{-1}\sim10^{6}$ yr \citep{Robson:2018svj}. This slow evolution allows for a Taylor expansion in the frequency:
\begin{equation}\label{eq:taylor_f}
	f_\text{gw}(t) = f_0 + \dot{f}_0t + \frac{1}{2}\ddot{f}_0t^2 + \mathcal{O}(t^3).
\end{equation}
The $t^0$ term, $f_0$, is referred to as the initial frequency or the `carrier frequency' in the following waveform description. 

The slow evolution of $f_\text{gw}$ also allows for a separation of timescales in a `fast-slow' decomposition. The fast timescale is the orbital timescale: $1/f\lesssim1$\,hr. The slow timescale is mostly governed, in addition to the weak effect of the frequency evolution of the source, by the motion of the {\it LISA} constellation in its heliocentric orbit. Therefore, the slow timescale is $\sim$1 yr. The large timescale difference was leveraged in \citet{Cornish:2007if} to create a fast frequency-domain waveform for GBs observed by {\it LISA}. The slow portion builds a sparsely sampled time-domain signal that accounts for the frequency evolution, as well as the projection of the signal onto the time-dependent arms of the {\it LISA} constellation. It is here where the ecliptic latitude, $\lambda$, and ecliptic longitude, $\beta_\text{sky}$, enter the computation. The sparse time-domain signal is then transformed to the frequency domain and convolved with a $\delta$ function at the carrier frequency. The convolution produces projections along the {\it LISA} constellation arms that are combined into time-delay interferometry \citep[TDI,][]{Tinto1999, Armstrong1999, Estabrook2000, Dhurandhar2002, Tinto2005, Vallisneri:2020otf, Tinto:2020fcc, Baghi:2020ygw, Page:2021asu} observables: $X,\ Y,\ Z$. These three TDI variables are correlated between them, but one could linearly combine them to build the noise orthogonal \cite{Vallisneri2005, aet}
\begin{align}
    A =& \frac{1}{\sqrt{2}}\left(Z-X\right) \qc\\
    E =& \frac{1}{\sqrt{6}}\left(X-2Y+Z\right) \qc \\
    T =&\frac{1}{\sqrt{3}}\left(X+Y+Z\right) \qs
\end{align}
These three TDI channels--$A,\ E,\ T$--are the final template waveforms that will go into the Bayesian analysis described in Section~\ref{sec:bayes}. 

\subsection{Inclusion of the third body in the waveform}

\begin{figure}
\begin{center}
\includegraphics[scale=0.3]{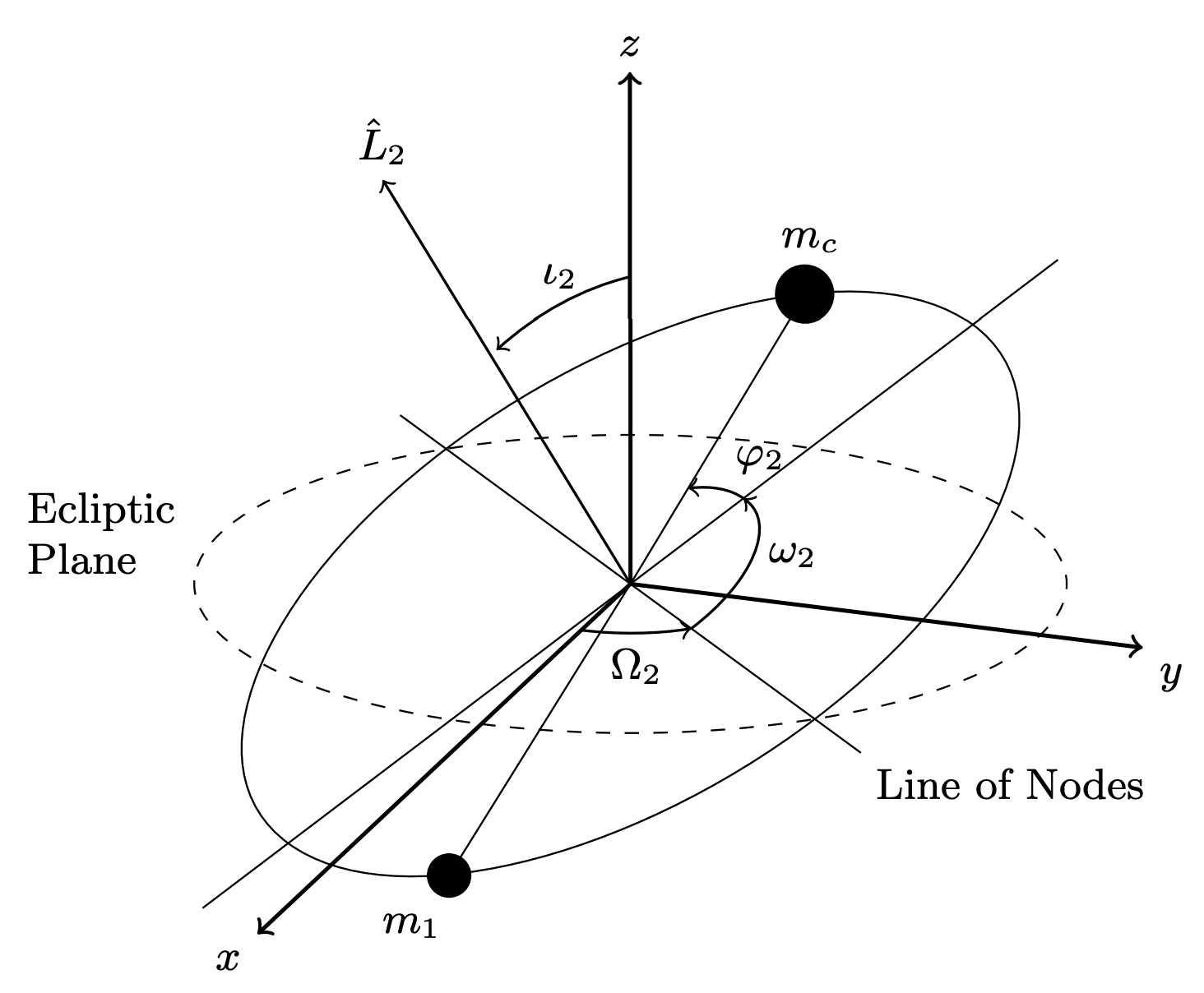}
\caption{A diagram illustrating the angular description of the third body ($m_c$) orbiting the inner GB ($m_1$). The orbit is described in the ecliptic reference frame. The orbital phase is $\varphi_2$. The three Euler rotation angles from the ecliptic plane are $-\omega_2$, $-\iota_2$, and $-\Omega_2$ \citep{Robson:2018svj}.}\label{fig:angular_diagram}
\end{center}
\end{figure}

The waveform as previously described includes only the contributions from GR to the shrinking of the two-body GB orbit. In order to include the effect of the third body on the waveform, we need only modify $\Psi_\text{gw}$. The adjustment must account for the acceleration of the center-of-mass of the inner binary caused by the third-body perturber. This information is included by red-shifting (or blue-shifting) the gravitational-wave frequency based on the line-of-site velocity, $v_{\parallel}$, of the inner binary:
\begin{equation}\label{eq:updated_psi}
	\Psi_\text{gw} = \phi_0 + 2\pi \int^t \left[1 + v_{\parallel}(t')\right]f_\text{gw}(t')dt' \qs
\end{equation}
The line-of-site velocity is given by \citet{Robson:2018svj}
\begin{equation}\label{eq:los_velocity}
	v_{\parallel}(t) = \mathcal{A}_2\left[\sin{\left(\varphi_2 + \varpi\right)} + e_2\sin{\left(\varpi\right)}\right] \qc
\end{equation}
where $\varphi_2$ is the orbital phase of the outer orbit and $e_2$ is the eccentricity of the outer orbit. The effective amplitude factor of the perturber, $\mathcal{A}_2$, is given by
\begin{equation}
	\mathcal{A}_2 = \frac{m_c}{m_2}\sqrt{\frac{m_2}{p_2}}\bar{A} \qc
\end{equation}
where $m_c$ is the mass of the perturber, $m_2$ is the total mass of the system, and $p_2$ is the semi-latus rectum of the outer orbit. $\bar{A}$ and $\varpi$ are based on the orientation of the third body orbit to the observer, where $\bar{A}$ is an effective amplitude and $\varpi$ is an effective phase. Figure~\ref{fig:angular_diagram} shows the orientation angles describing the orbit of the perturber involved in the calculation of $\bar{A}$ and $\varpi$. The three Euler rotation angles from the ecliptic plane are $-\omega_2$, $-\iota_2$, and $-\Omega_2$. With $\theta=\pi/2-\beta_\text{sky}$ and $\phi=\lambda$ as the polar and azimuthal coordinates of the sky location, respectively, we define $C(\theta, \iota_2, \phi, \Omega_2)= \cos{\theta}\sin{\iota_2} + \sin{\theta}\cos{\iota_2}\sin{\left(\phi - \Omega_2\right)}$ and $S(\theta, \phi, \Omega_2) = \sin{\theta}\cos{\left(\phi - \Omega_2\right)}$.
The effective amplitude is then $\bar{A}^2 = C^2 + S^2$ and the effective phase is $\varpi = \omega_2 + \bar{\phi}$, where $\tan{\bar{\phi}} = \frac{C}{-S}$. Due to degeneracies between the parameters needed to determine $\mathcal{A}_2$, it is not possible to gain independent information on all of them. Therefore, the sampling algorithm described in Section~\ref{sec:bayes} will compress all of these angles and only sample in $\mathcal{A}_2$.  For more detailed information on these constructions see \citet{Robson:2018svj}.

The line-of-sight velocity is determined from Eq.~\eqref{eq:los_velocity} and the inversion of Kepler's equation. The phase contribution from the line-of-sight velocity ($v_\parallel$ term in Eq.~\eqref{eq:updated_psi}) is integrated numerically for each system. Within the fast GB waveform generation methodology, the effect of the line-of-sight velocity on the waveform evolves on a timescale of $P_2\sim1$yr, which means it evolves on a `slow' timescale compared to the orbital timescale of the inner binary. This slow phase change is directly incorporated into the determination of $\Psi_\text{gw}$ in addition to the frequency evolution of the source and the effect of the {\it LISA} orbital motion. With the addition of the phase correction due to the perturber, the acceptable sparsity of the slow time-domain part of the waveform now depends on the value of $P_2$. For values of $P_2$ that are much shorter than $\sim 1$yr, the number of samples is increased in order to properly sample the third-body orbit. 

\section{Bayesian analysis with {\it LISA}}\label{sec:bayes}

Gravitational-wave data analysis is centered around extracting information on a signal, $s(t)$, that is submerged in noise, $n(t)$, with the full datastream, $d(t)$, consisting of the sum of these two components: $d(t) = s(t) + n(t)$. To perform this extraction, templates, $h(t)$, are designed to match $s(t)$ to the greatest degree possible. In our work, $h$ will be generated directly in the frequency domain with the fast waveform generator discussed in Section~\ref{sec:gbwave}.

The statistical analysis is governed by Bayes' theorem:
\begin{equation}\label{eq:Bayes}
	p(\vec{\Theta}|d, \Lambda) = \frac{p(d|\vec{\Theta}, \Lambda)p(\vec{\Theta}|\Lambda)}{p(d|\Lambda)} \qc
\end{equation}
where $\vec{\Theta}$ is the parameterization of model $\Lambda$. In this work, there are two different models: the base template and third-body template. The posterior probability on the parameters of the source is $p(\vec{\Theta}|d, \Lambda)$. The prior density, $p(\vec{\Theta}|\Lambda)$, represents the incorporation of prior knowledge into the posterior distribution on the parameters. The Likelihood, $\mathcal{L}(\vec{\Theta}|\Lambda)=p(d|\vec{\Theta}, \Lambda)$, is the gravitational-wave specific quantity that will be computed in the process of determining the posterior distribution. It represents the probability that the observed datastream could be produced by a source with parameters $\vec{\Theta}$ under model assumption $\Lambda$. The denominator on the right side of Eq.~\eqref{eq:Bayes} is the Evidence, which is an integral of the numerator over all of parameter space: $Z(\Lambda)=p(d|\Lambda) = \int_{\vec{\Theta}}\mathcal{L}(\vec{\Theta}|\Lambda)p(\vec{\Theta}|\Lambda)d\vec{\Theta}$. This term is generally intractable to compute exactly in the gravitational-wave case. However, there are approximate methods to compute this term that will be discussed below. 

The gravitational-wave Likelihood has the shape of the Gaussian distribution on the datastream residual after the template is applied:
\begin{equation}\label{eq:like}
	\ln{\mathcal{L}} \propto -\frac{1}{2}\langle d-h | d-h  \rangle \qc
\end{equation}
where $\langle a | b \rangle$ is the noise-weighted inner product between two time-domain datastreams, $a(t)$ and $b(t)$. In the analysis performed for this work, Eq.~\eqref{eq:like} is factored: \mbox{$\langle d-h | d-h \rangle = \langle d | d \rangle + \langle h | h \rangle - 2\langle d | h \rangle$}. From these terms, we get the optimal S/N achievable by template $h$: $\sqrt{\langle h|h \rangle}$, and the extracted S/N of a template against the datastream: $\langle d | h \rangle / \sqrt{\langle h|h \rangle}$. Assuming the noise is stationary and Gaussian, we can write down the integral that gives the noise-weighted inner product:
\begin{equation}\label{eq:inner}
	\langle a | b \rangle = 4\text{Re}\int_0^\infty \frac{\tilde{a}(f)^* \tilde{b}(f)}{S_n(f)} df \qc
\end{equation}
where $\tilde{a}(f)$ is the Fourier transform of \mbox{$a(t)$: $\tilde{a}(f) = \mathcal{F}\{ a(t) \}$}; and $S_n(f)$ is the one-sided power spectral density (PSD) in the noise. For the noise PSD, we use the `SciRDv1' noise curve from the {\it LISA} Data Challenges Working Group \citep{SciRD1}. Please note the inner product is really a sum over the inner products in channels $A$ and $E$. We do not include the $T$ channel in our analysis because it is not sensitive to the signals examined here under the assumed equal-arm length configuration to the orbit of the LISA constellation. In Eq.~\eqref{eq:inner}, the inner product can be maximized over the initial phase parameter by taken the modulus of the complex integral rather than taking just its real value. We will use maximization over initial phase to make a first cut of undetectable systems. This will be discussed further in Section~\ref{sec:catalogue_analysis}.

\subsection{Search and Parameter Estimation with MCMC}

Throughout our analysis we will use Markov Chain Monte Carlo (MCMC) sampling techniques to draw samples from the posterior distribution. Our MCMC sampler ({\sc Eryn}; \citeauthor{karnesis-in-prep}) is based on {\sc emcee} \citep{emcee} and uses the parallel tempering scheme from {\sc ptemcee} \citep{Vousden2016}. The proposal we use is the Stretch proposal from \citet{Goodman2010}.  In Section~\ref{sec:locating_max_ll}, we use these techniques to search for the maximum Likelihood of a base template driven only by GR compared against an injection waveform containing the effect of a third body. After locating maximum Likelihood values, we use MCMC runs to generate full posterior distributions for computing the Evidence via Thermodynamic Integration. The log of the Evidence (relative to the prior) at $T=1$ is given by \citep{Goggans2004ThermoInt, Lartillot2006ThermoInt}
\begin{equation}\label{eq:evidence_integral}
	\log{Z(1)}  =  \log{Z(0)} + \int_0^1 d\beta \left \langle \log{\mathcal{L}}\right\rangle_\beta\qc
\end{equation}
where $\beta=1/T$ and $\left \langle \log{\mathcal{L}}\right\rangle_\beta$ is the average of the log-Likelihood at inverse temperature $\beta$. With a normalized prior distribution $\log{Z(0)}=0$. In practice, this integral is approximated with a sum across discrete rungs of a temperature ladder between $\beta=0$ and $\beta=1$. Then, the Bayes Factor is simply the ratio of the Evidence between two models: $B_{12} = Z_1/Z_2$. A Bayes Factor of less (more) than 1 indicates model 2 (1) is favored. Here,  we will compute the log of the Bayes Factor. Table~\ref{tb:bayesfactor} gives basic `confidence' levels for different ranges of $B_{12}$ and $2\log{B_{12}}$.

\begin{table}
\begin{center}
\begin{tabular}{| c | c | c |}
\hline
 $B_{12}$ & $2\log{B_{12}}$ & Evidence for model 1 \\ 
\hline
 <1 & <0 & Negative (supports model 2) \\  

1 to 3 & 0 to 2 & Not worth more than a bare mention \\

3 to 12 & 2 to 5 & Positive \\

12 to 150 & 5 to 10 & Strong \\

>150 & >10 & Very Strong \\
\hline
\end{tabular}
\caption{`Confidence' levels based on the Bayes Factor, $B_{12}$ \citep{Raftery1996,Cornish:2007if}. We consider $2\log{B_{12}}\geq5$ to indicate a detectable source.}\label{tb:bayesfactor}
\end{center}
\end{table}

The maximum Likelihood, posterior distribution, and Evidence are determined by using MCMC algorithms with an ensemble of $n_w=26$ walkers in each of $n_T=1000$ temperatures (26,000 total walkers). The waveform and Likelihood codes are accelerated with Graphics Processing Units (GPU), which allows for a large number of temperatures to be evaluated to better determine the temperature-Likelihood profile to be integrated. The target distribution is examined with the cold chain at a temperature of 1. The highest temperature is set to infinity to probe the prior distribution. 100 inverse temperatures ($\beta=1/T$) are log-spaced between $10^{-10}$ to $10^{-4}$. This set spans the portion of the temperature ladder where the signal is entirely suppressed by the tempering (S/N$/\sqrt{T}\sim1$). The other 900 inverse temperatures are log-spaced from $10^{-4}$ to $10^{0}$ in order to resolve with high density the turnover where the signal is no longer suppressed. This setup also captures in great detail all temperatures where the signal can be found. 

All prior distributions on each parameter are treated independently. The amplitude ($A$) prior is log-uniform between $10^{-24}$ and $10^{-20}$. The initial frequency ($f_0$) prior is uniform between 0.5 mHz and 20 mHz. The priors on the $\phi_0$ and $\lambda$ are uniform from zero to 2$\pi$, while $\psi$ is uniform from zero to $\pi$. The inclination angle is uniform in $\cos{\iota_1}$. The ecliptic latitude is uniform in $\sin{\beta_\text{sky}}$. The frequency derivative prior is determined separately for each system so that its range spans the derivative values associated with a range of chirp masses at each system's true initial frequency. The chirp mass range is $10^{-3}\ \Msun$ to 1.0 $\Msun$, in agreement with the simulated population (cf. Section~\ref{sec:dwd_pop}).

The third-body angular parameter $\varpi$ is uniform from zero to $2\pi$. The time of periastron passage ($T_2$) prior is uniform from zero to $P_2$ for each specific system. The eccentricity prior was allowed to vary uniformly from zero to 0.985. The $P_2$ prior was also uniform. Its lower bound was 0.1 times the injection value. If $P_2 < T_\text{obs}/2$ with $T_\text{obs}=4$yr, the upper bound was chosen to be 2.2 yr. If the third-body period was between 2 yr and 3.2 yr, the maximum allowable $P_2$ value was 32 yr. If the injection value was greater than 3.2 yr, the upper bound on $P_2$ was set to $10$ times the injection value. The choice of prior on $A_2$ is complicated because it depends on a large number of input parameters, some of which are not accessible in the extraction process. Therefore, we chose an expansive $A_2$ prior given by the one-dimensional marginalized distribution on $A_2$ from the injection catalogue. This distribution was determined using {\sc kalepy} \citep{Kelley2021}. When the $P_2$ prior limits were adjusted, the $A_2$ distribution was also changed to reflect the new period bounds. These prior choices were verified throughout the process to ensure they did not affect the results.




\section{CBP and BD Catalogue Analysis}\label{sec:catalogue_analysis}

\subsection{Locating the maximum Likelihood point}\label{sec:locating_max_ll}

The determination of the detectability of the sub-stellar objects in the catalogue is done through various steps.
The first step is to compute the phase-maximized Likelihood between the third-body injection waveform ($d$) and the GR-only template waveform ($h$). This computation is performed at parameters of the base template that are exactly equal to the inner binary parameters of the third-body injection (except for $\phi_0$ due to the phase maximization). If the phase-maximized Likelihood for a given source in the population is greater than $-2$ (and, by definition, less than zero), we consider this third body undetectable. The value of $-2$ was chosen conservatively based on initial investigations of the Evidence-based detectability study described below. We also eliminate sources from consideration if their S/N is less than 10. These two cuts eliminate 8349/12737 [66\%] sources from the original catalogue (6278/10287 [61\%] BDs and 2071/2450 [85\%] CBPs).

Subsequently, the remaining sources are tested with stochastic sampling via MCMC to further refine the determination of the maximum Likelihood by allowing all parameters to vary throughout the prior volume. This is a necessary operation because, in general, when the template model does not match the injection model, the maximum Likelihood point will not be at the exact injection parameters. This can be viewed effectively as a `burn-in' or `search' phase: the code searches for combinations of parameters to best mimic the effect on the waveform created by the third-body orbit.   The search is run until it converges to a maximum Likelihood value. If this value is between -2 and zero, the associated third-body source is considered undetectable. For many sources in the catalogue, but not all, the search does locate a maximum Likelihood that is noticeably higher than the original phase-maximized Likelihood determined in the first step described above. These sources require a bias against the injection parameters in order for the base template to match the effect of the perturber. 

\begin{figure}
\begin{center}
\includegraphics[trim=0.cm 2.8cm 0.5cm 0.5cm, clip,width=\columnwidth]{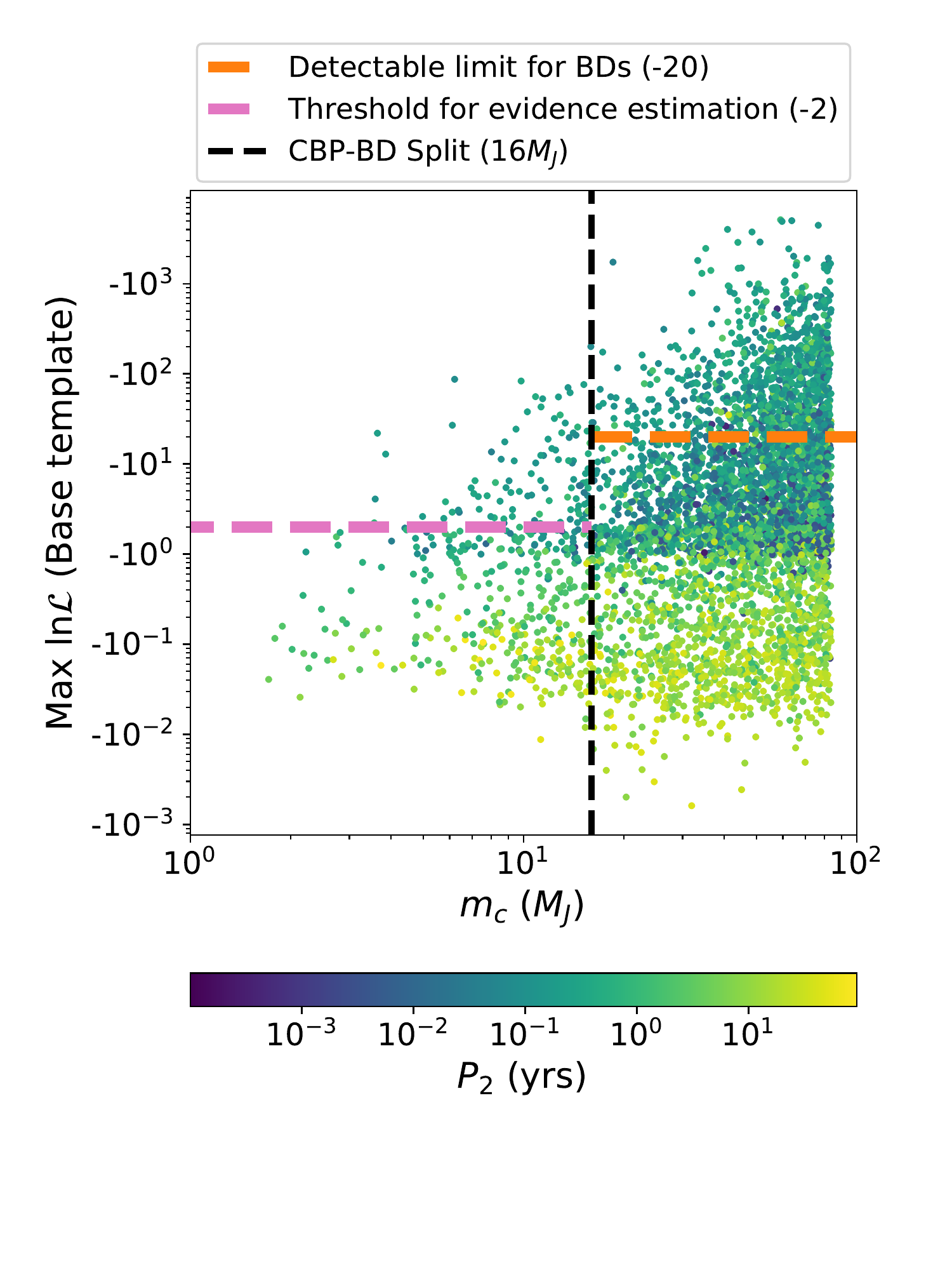}
\caption{Output from the search for Likelihood maximization using MCMC techniques. All sources shown have $\ln{\mathcal{L}}<-2$ from the initial template comparison between the base template and the third-body template at the injection parameters. The scatter points show, vertically, the $\ln{\mathcal{L}_\text{max}}$ achieved after letting the parameters for the base template vary across the prior domain. The horizontal axis gives the true value of the perturber's mass. Our chosen mass separation between CBPs and BDs is displayed with the black dashed line ($m_c=16\ \Mj$). For CBP systems ($m_c<16 \Mj$), a conservative $\ln{\mathcal{L}}_\text{max}$ cut of less than -2 (magenta dashed line) was applied to choose systems for accurate Evidence estimation due to the computation's high cost. After applying the accurate Evidence estimation to all CBP systems with $\ln{\mathcal{L}}_\text{max}<-2$, a conservative cut of $\ln{\mathcal{L}}_\text{max}<-20$ (orange dashed line) was determined as a reliable detection cut for determining detectable BD systems.}\label{fig:likelihood_vs_mass_scatter}
\end{center}
\end{figure}

Figure~\ref{fig:likelihood_vs_mass_scatter} shows the maximum log-Likelihood found for each system that passed the first cut with the phase-maximized Likelihood at the injection parameters. These are shown for both CBP and BD systems with applicable cuts for each population shown with dashed lines. The mass of the third body is shown on the horizontal axis and the period of the third body is indicated by color according to the color bar. The separation in SSO period $P_2$ is stark and will be discussed further below. It is also clear the importance of having a larger third-body mass in creating a strong enough waveform perturbation for detection. 

This Likelihood maximization process removes 1982/4388 [45\%] of the remaining sources (272/379 [72\%] CBPs and 1710/4009 [43\%] BDs), leaving 107 CBPs and 2299 BDs.  

\subsection{Estimating the Evidence}\label{sec:estimate_evidence}

\begin{figure*}
\begin{center}
\includegraphics[trim=1.cm 4.8cm 6.5cm 2cm, clip,width=\textwidth]{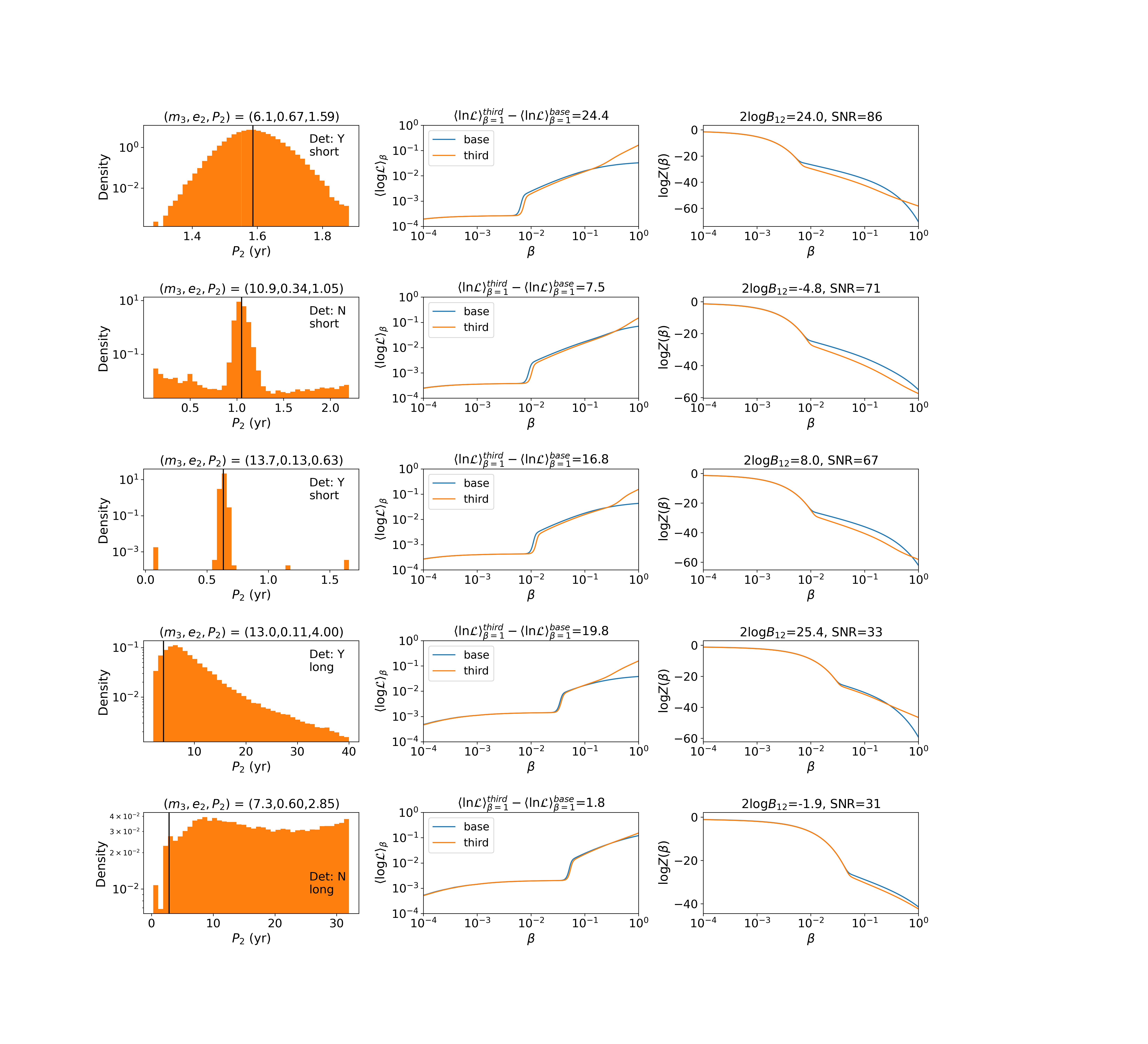}
\caption{Five examples of results from the accurate Evidence estimation analysis. The first column is the one-dimensional marginalized histogram on the third-body orbital period ($P_2$). Please note the vertical axis is log-scaled. At the top of the $P_2$ histograms are the mass ($\Mj$), eccentricity, and true period (yr) of the pertuber in each row. The detectability is also given as (Y)es or (N)o, followed by whether the third-body period is in the short- or long-period grouping. The second column shows the average log-Likelihood at each inverse temperature; it is this curve that is integrated to calculate the Evidence. The base template results are shown in blue and the third-body results are shown in orange. The title in this column lists the difference in the cold-chain ($\beta=1$) average log-Likelihoods between the two templates. The final column illustrates the thermodynamic integral as a function of the inverse temperature. The Evidence-related curves are analyzed in detail in Section~\ref{sec:estimate_evidence}. The final column titles display $2\log{B_{12}}$ and the S/N for each source.
}
\label{fig:final_period_evidence_full}
\end{center}
\end{figure*}

While full Evidence computations of the large number of remaining BDs would be computationally intensive, the smaller number of remaining CBPs allows us to properly estimate the Evidence ratio for each system using Thermodynamic Integration.
For this reason, we chose to focus on the posterior properties of the CBP population. We then address the detections of BDs by determining a conservative maximum log-Likelihood difference that encapsulates all detectable systems according to the detailed Evidence computations of the CBP sources. This allows us to use the results from the last section to make a conservative detectability cut for the BD systems based on the maximum log-Likelihood of the GR-only template. 

Figure~\ref{fig:final_period_evidence_full} shows five CBP sources from top to bottom that aid in clarifying the various properties of the Evidence computations. These individual sources and their properties will be discussed shortly. The one-dimensional marginalized histogram of the outer period $P_2$ is shown for each source in the left column. The centre column compares the average log-Likelihood over inverse temperature between the two models. The Evidence integral (Eq.~\ref{eq:evidence_integral}) as a function of inverse temperature is shown in the right column. 

The behavior of the average log-Likelihood over inverse temperature (centre column) is similar for all systems tested. At low $\beta$, where the signal is suppressed, the log-Likelihoods are effectively equivalent. In the transition range, where the signal is found, there is some unexpected and interesting behavior. The signal suppression turnover point generally follows proportionally to the S/N of the source. Given the base template is inherently incorrect and the third-body template has no modelling error compared to the injection, it is expected that the average S/N of a given temperature rung will be higher for the third-body template. As can be seen in the centre column of Figure~\ref{fig:final_period_evidence_full}, this is consistently not the case: the base template finds the signal at a slightly lower inverse temperature compared to the third-body template. We believe this is a manifestation of the five-dimension penalty associated with the addition of the third-body parameters. As the inverse temperature moves above this cross-over point, the average log-Likelihood of the third-body template remains less than the base template for a stretch of inverse temperature values. Then, for higher inverse temperatures, the average log-Likelihood for the third-body template increases and crosses above the base template. Between these two cross-over points, the Evidence integral difference accumulates a negative value indicating favorability of the base model. This log-Evidence-difference penalty is roughly -2 to -4 with shorter periods incurring a more negative penalty, which can be observed in the visual width between the two curves at the turnover point. This penalty can be more negative for some systems with shorter periods. The sources that are detectable have a high enough log-Likelihood difference between the templates in the cold chain ($\beta=1$) that the difference integral accrues enough positive effect on the Evidence ratio to overcome the dimensionality penalty. 

Beyond the general behavior of the Evidence integral, separate consideration must be given towards the systems with $P_2<T_\text{obs} / 2$ versus systems with $P_2>T_\text{obs} / 2$, which we will refer to as short- and long-period systems, respectively. This middle point represents the Nyquist criterion for proper sampling of the full third-body orbit: at $P_2<T_\text{obs} / 2$ the orbital motion of the perturber is sampled well enough to resolve its frequency (or period), generally preserving a more Gaussian structure on the $P_2$ one-dimensional marginalized posterior if the third body is detectable. The short-period binaries provide a strong indication confirming the accuracy of the Evidence computations: short-period perturbers that are \textit{detectable} are represented by a $P_2$ histogram that is a tight Gaussian dropping off until it reaches 10 or fewer samples in a bin on both sides. The first row in Figure~\ref{fig:final_period_evidence_full} is an example of a $P_2$ histogram for a detectable short-period source. The histogram shows the log base 10 of the density to show in detail the samples near the edge of the distribution's tails. Notice it does not extend throughout the prior range. Additionally, in some of these $P_2$ distributions, there is a slight skew towards larger periods. 

An undetectable short-period third body is shown in the second row of Figure~\ref{fig:final_period_evidence_full}. There is a roughly Gaussian distribution around the true value; however, the Gaussian falls off into a background created by the prior distribution, which is not easily seen with linear density on the vertical axis but can be seen clearly with the log scale. In some examples of undetectable short-period sources, artifacts can also be observed at yr/2 and 1 yr due to confusion related to the orbit of the {\it LISA} constellation. 

Some detectable sources with periods less than $T_\text{obs} / 4=1$ yr will show a tight Gaussian around the true parameter, with a very small percentage of samples scattered throughout the prior range. This is shown in the third row of Figure~\ref{fig:final_period_evidence_full}. There is a key difference qualitatively indicating this source is detectable compared to sources like those shown in the second row: the small number of samples found away from the true peak are confined to only a small number of other bins clearly showing there is no continuity in the potential background prior samples. All undetectable sources yield posterior distributions that are non-zero across the prior range. 

With $P_2>T_\text{obs} / 2$, the ability to resolve the period of the pertuber, as well as its Gaussianity in its posterior, is diminished. Representative examples of a detectable and an undetectable long-period source are shown in rows four and five of Figure~\ref{fig:final_period_evidence_full}, respectively. The $P_2$ histogram of the detectable long-period source is heavily skewed to longer periods filling the entire prior range. It peaks at longer periods than the true value, which is observed in sources with $P_2\geq T_\text{obs}$. For $T_\text{obs}/2 < P_2 \lesssim T_\text{obs}$, the distribution peaks at the true value. In long-period cases, where the Nyquist criterion on the third-body orbit is not met, every source is effectively consistent with longer periods until the effect diminishes and the right side of the posterior reaches approximately zero. 

Undetectable long-period sources (e.g., row five in Figure~\ref{fig:final_period_evidence_full}) show two distinct properties compared to detectable sources. The first difference is that the distribution does not tend towards lower counts at longer periods; rather, the histograms tend to turn upward at the end due to the upper boundary on the uniform prior, indicating the signal is not matched well at the large period end of the prior. The second difference, which is specific to sources with $T_\text{obs}/2 < P_2 \lesssim T_\text{obs}$, is a low-density region appearing at $P_2/2$, which can be seen to the left of the histogram in the fifth row. 

The detectable and undetectable short- and long-period events segregate into a few roughly distinct Evidence groups. The short period undetectable binaries are usually found at the lowest Evidence values due to the larger penalty at the S/N-temperature turnover point. Long-period undetectable sources populate the Evidence spectrum from $-5\lesssim2\log{B_{12}}\lesssim0$, with marginally detectable long-period sources having Evidence values of less than 5 and greater than zero. All of the detectable sources with $2\log{B_{12}}>5$ congregate together. The lack of short-period sources found between $-5\lesssim 2\log{B_{12}} \lesssim 5$ once again highlights the Nyquist criterion on the third-body orbit: sources with $P_2 < T_\text{obs}/2$ are mostly either strongly detectable or strongly undetectable due to the clearer resolution of their orbital frequency.  

\subsection{Detectable population}

\begin{figure}
\begin{center}
\includegraphics[trim=0.cm 0cm 1.cm 0.5cm, clip,width=\columnwidth]{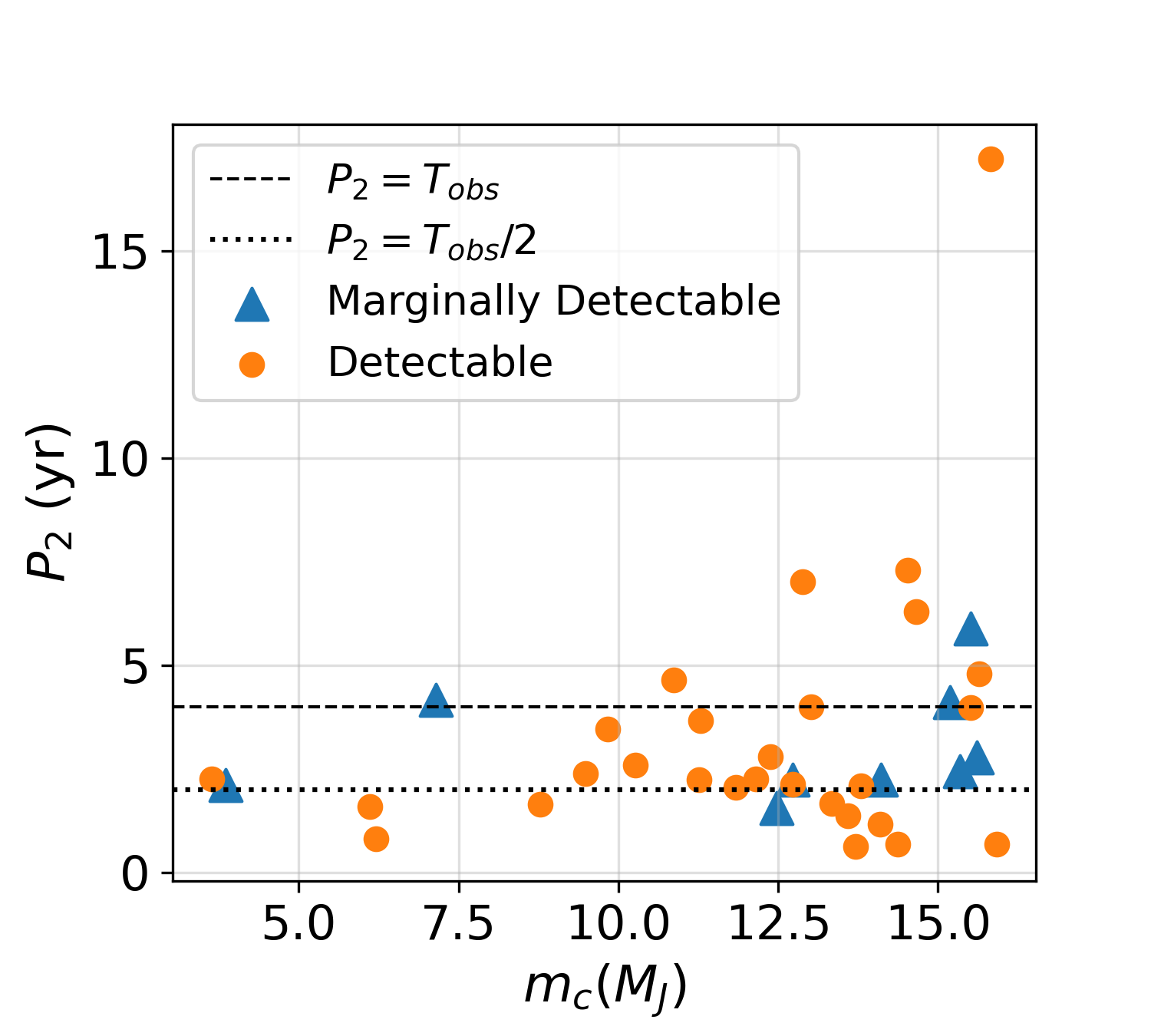}
\caption{Scatter plot showing the third-body mass (horizontal axis) and period (vertical axis) of detectable (orange dots) and marginally detectable (blue triangles) CBP systems. It must be noted, these are the injection values, not recovered. Horizontal lines correspond to $T_\text{obs}$ and $T_\text{obs}/2$ as indicated in the legend.}

\label{fig:detectable_planets_scatter}
\end{center}
\end{figure}

We will first examine the detectable CBP population with each source's Evidence ratio estimate as the detection metric. We make a conservative cut at $2\log{B_{12}}> 5$, indicating strong favorability of the third-body template. We find that there are 28/2450 [1\%] detectable CBP systems within our modeled catalogue. The planetary mass and orbital periods of the detected and marginally detected ($0<2\log{B_{12}}<5$) sources are shown in Figure~\ref{fig:detectable_planets_scatter}. Most detections are found with $P_2 < T_\text{obs}$ with more detections at larger mass, as would be expected since increased mass strengthens the effect of the perturber. The lowest mass detected is $\sim4\ \Mj$. We remind the reader that the population detected is a function of the orbital architecture of the injected SSO population, which was built through a single stochastic draw from the various orbital parameter distributions (Sec. \ref{sec:ssoinjection}). To robustly determine the range of detectable masses, multiple injected populations must be analysed. Such work is currently underway and will be presented in an upcoming paper (\citeauthor{Danielski-in-prep}).

Following the Evidence computations on the CBP systems, and given that all CBP sources with $\Delta\ln{\mathcal{L}}>20$ had detectable Evidence ratios, we decided that, for BD systems, a cut of --20 on the maximum log-Likelihood achieved by the base template is sufficient for likely detection. This value was spot-checked with a smaller sample of BD systems. With this conservative cut, 951/10287 [9\%] BD systems are detected. This cut probably removes a few hundred detectable sources.

\begin{figure}
\begin{center}
\includegraphics[trim=0.cm 1.6cm 0.5cm 2.5cm, clip,scale=0.7]{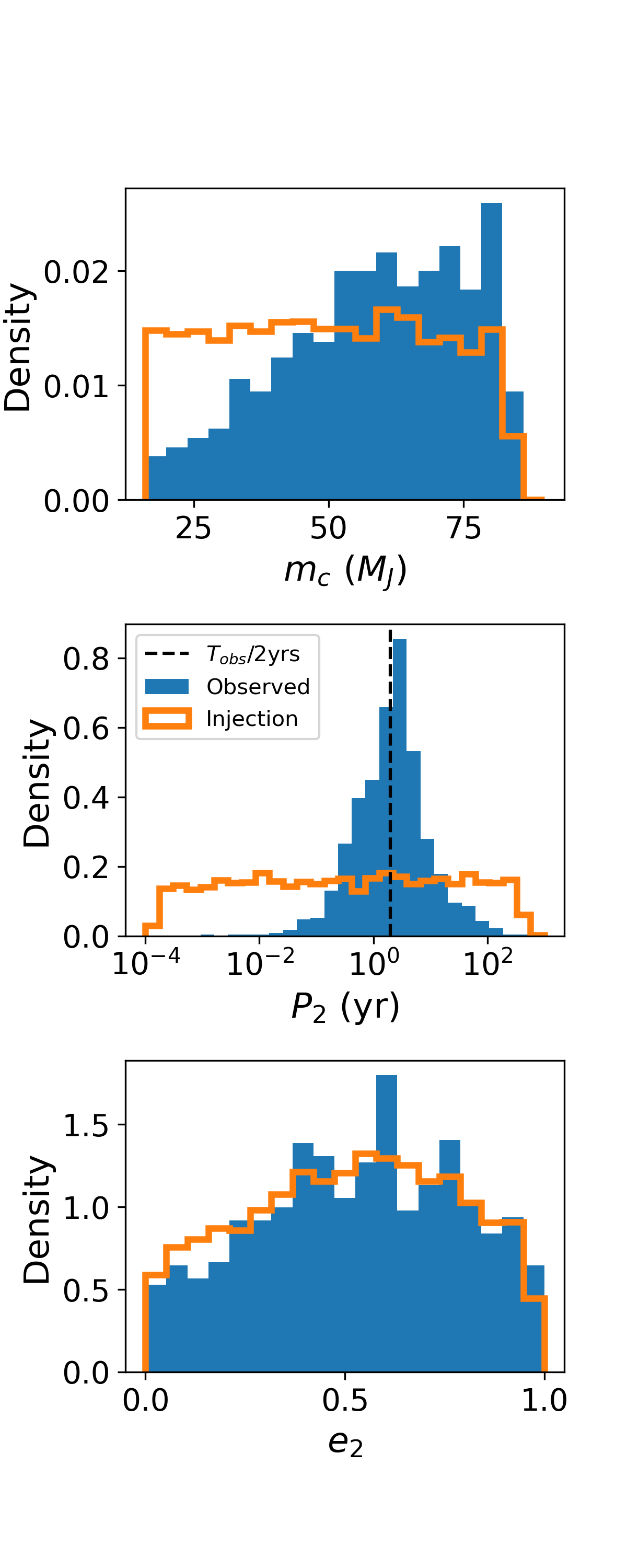}
\caption{Histograms showing the injection parameters of BD systems around inner GBs. The mass ($m_c$), period ($P_2$), and eccentricity ($e_2$) of the third-body are shown from top to bottom, respectively. The injected population is shown in orange. The detectable population is shown in blue. Future work is needed to understand how these histograms vary over different population models, as well as when examining recovered parameters, rather than injected.}\label{fig:output_bds_hist} 
\end{center}
\end{figure}

The mass, period, and eccentricity of the detectable BD population is shown in Figure~\ref{fig:output_bds_hist}. Please note these are the histograms of the injected values that are detected, not observed values. As expected, the highest number of detections is seen at higher mass and the period histogram peaks at $P_2 < T_\text{obs}$, confirming the trends presented in \cite{Tamanini:2018cqb} and \cite{Danielski:2019rvt}. The eccentricity distribution of observed sources does strongly resemble the injected population. Initial checks on BD systems show weaker constrains on the eccentricity (which matches output from the CBP analysis), but a deeper examination is required at a population level. This is a topic for future work (\citeauthor{Danielski-in-prep}). 

\subsection{CBP Posterior Analysis}\label{sec:cbp_post}

\begin{figure*}
\begin{center}
\includegraphics[trim=0.cm 0.5cm 0cm 0.5cm, clip,width=\textwidth]{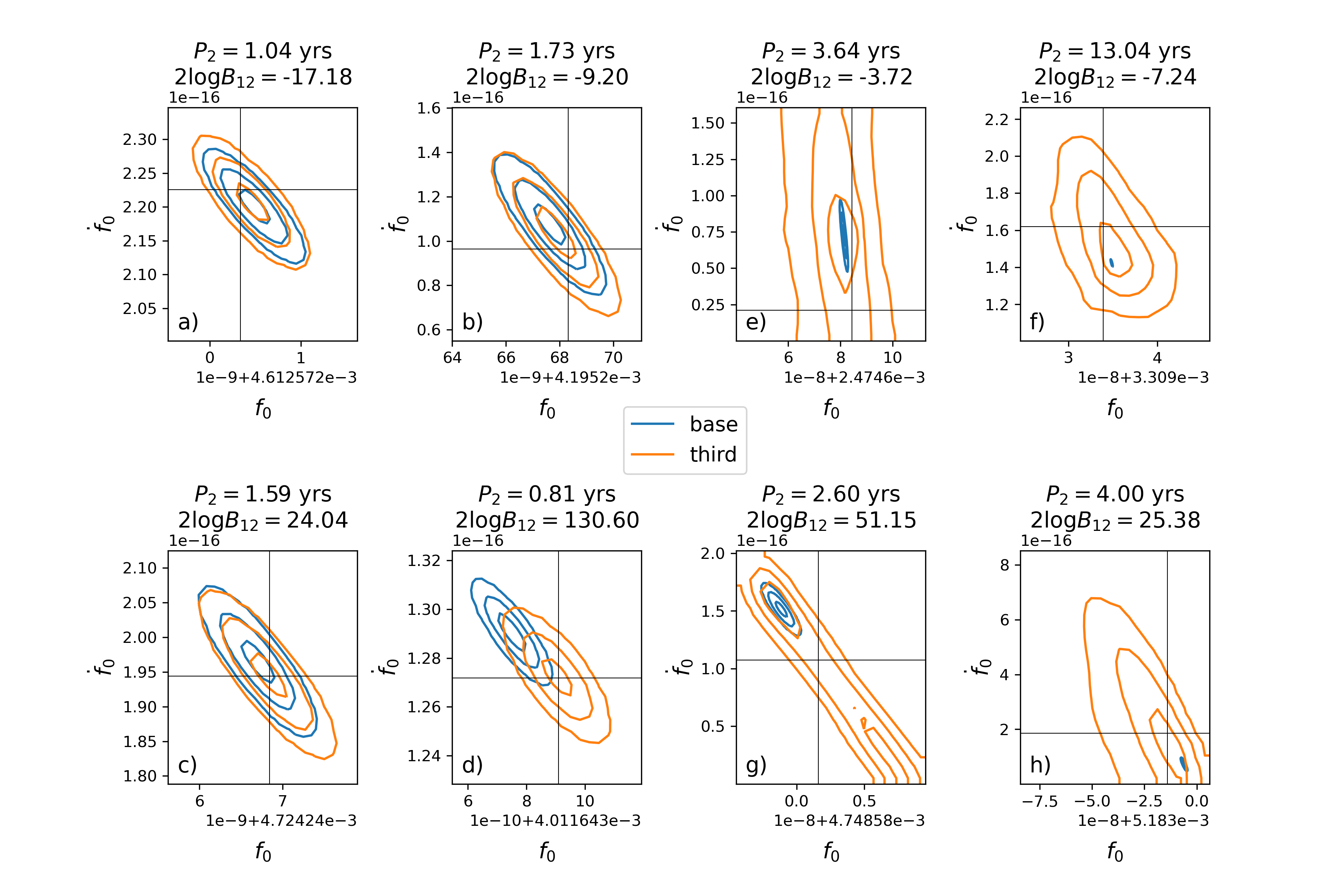}
\caption{The two-dimensional marginalized posterior distributions ($1\sigma,2\sigma,3\sigma$ contours) in the $f_0-\dot{f}_0$ plane for eight CBP systems in our catalogue. The distribution for the base GR-only template is shown in blue and the distribution including the third-body effect is shown in orange. The true injection parameters are shown with the horizontal and vertical thin black lines. The injection always consists of the third-body inclusive template. The upper and lower rows display undetectable ($\log{B_{12}}<5$) and detectable ($\log{B_{12}}\geq5$) systems, respectively. The left two columns are sources with $P_2<T_\text{obs}/2$ ($T_\text{obs}=4$ yrs), which represents the Nyquist sampling criterion to completely sample the third-body orbital evolution. The right two columns are third-body objects with periods longer than $T_\text{obs}/2$. Twice the log of the Bayes factor and the third-body period are shown above each individual plot. Undetectable sources lead to a third-body posterior distribution that usually centers on the mean of the base template distribution, indicating a bias on recovered parameters will occur whether or not the correct template is used. Detectable short-period sources have a roughly Gaussian character to their posterior distributions in $f_0-\dot{f}_0$. On the contrary, detectable sources with periods longer than half the observation time have wider and more unpredictable shapes.}\label{fig:f_fdot}
\end{center}
\end{figure*}

The detailed Evidence computations for the CBP systems also produced posterior distributions in the cold chains. Here, we will focus on examining patterns observed in important CBP system parameters across the population by looking at two-dimensional marginalized posterior distributions. Full posterior distributions are available upon request to the authors.

Figure~\ref{fig:f_fdot} shows the relation between the frequency and the frequency derivative over eight example systems. The first thing to note is that the maximum log-Likelihood locating operation from Section~\ref{sec:locating_max_ll} preferentially preserves sources with higher frequency and frequency derivative on the inner binary. This is due to the increased inner binary frequency content in the chirping signal that must contend with the doppler shifting from the outer perturber. Higher $f_0-\dot{f}_0$ sources without any third body tend to have posterior distributions that are multivariate correlated Gaussians. The Gaussian behavior is still preserved in the posterior distributions on the base GR-only template, even while the distributions are biased away from the true parameters. When examining the bias of the base template, it must be noted that, in general, the base template holds a stronger constraint on the frequency and frequency derivative compared to the third-body template due to the lower dimensionality and lack of the doppler shifting effect of the perturbing body. 

The behavior of the third-body template can once again be separated into groups with short ($P_2 < T_\text{obs}/2$) and long ($P_2 > T_\text{obs}/2$) periods. The short-period systems ($a,b,c$, and $d$ in Figure~\ref{fig:f_fdot}) all have multivariate Gaussians for their posteriors. Detectable short-period sources ($c$ and $d$ in Figure~\ref{fig:f_fdot}) show a third-body template distribution that is centered on the true point. The base template distribution shows a slight bias from the true point, with the relative magnitude of the bias proportional to the Evidence ratio. 

Undetectable short-period sources ($a$ and $b$ in Figure~\ref{fig:f_fdot}) have distributions that are in general not centered on the true parameters, but centered on, or near, the base template distribution. In these cases, the base template distribution, which inherently exists in a lower dimensional space, pulls the posterior weight away from the higher dimensional third-body template. This can also be directly observed in the maximum Likelihoods found in the third-body distributions: even if an MCMC walker was started near the maximum Likelihood point with a log-Likelihood that was approximately zero, over the course of burn-in the walkers would all drift towards the base template distribution leaving the maximum log-Likelihood in the final MCMC chains to be away from zero and closer to the maximum log-Likelihood of the base template distribution. This indicates that, whether we use the correct template model or not, resulting samples will be biased compared to the true parameters.

The long-period sources, which are shown as plots $e,f,g$, and $h$ in Figure~\ref{fig:f_fdot}, have third-body template behavior that is no longer Gaussian, once again due to the failure to meet the Nyquist criterion on the sampling of the third-body orbit. The result of this is much wider posteriors with less predictable qualities that are stretched due to the slow doppler shifting caused by the perturber. Undetectable sources ($e$ and $f$) tend to peak at the center of the base template distribution. With longer period sources, this is also true for some detectable systems, like the one shown as plot $g$ in Figure~\ref{fig:f_fdot}. This example shows a very unique posterior that has 1$\sigma$ contours surrounding both the base template distribution and a point that is roughly opposite from the base template distribution across the true point. The true point is contained in the 2$\sigma$ contour in this case. This once again leads to an issue where either template will create a bias on the extracted parameters, but, in this case, with larger periods, this is even true for detectable third-body systems. 

\begin{figure*}
\begin{center}
\includegraphics[trim=0.cm 0.5cm 0cm 0.5cm, clip,width=\textwidth]{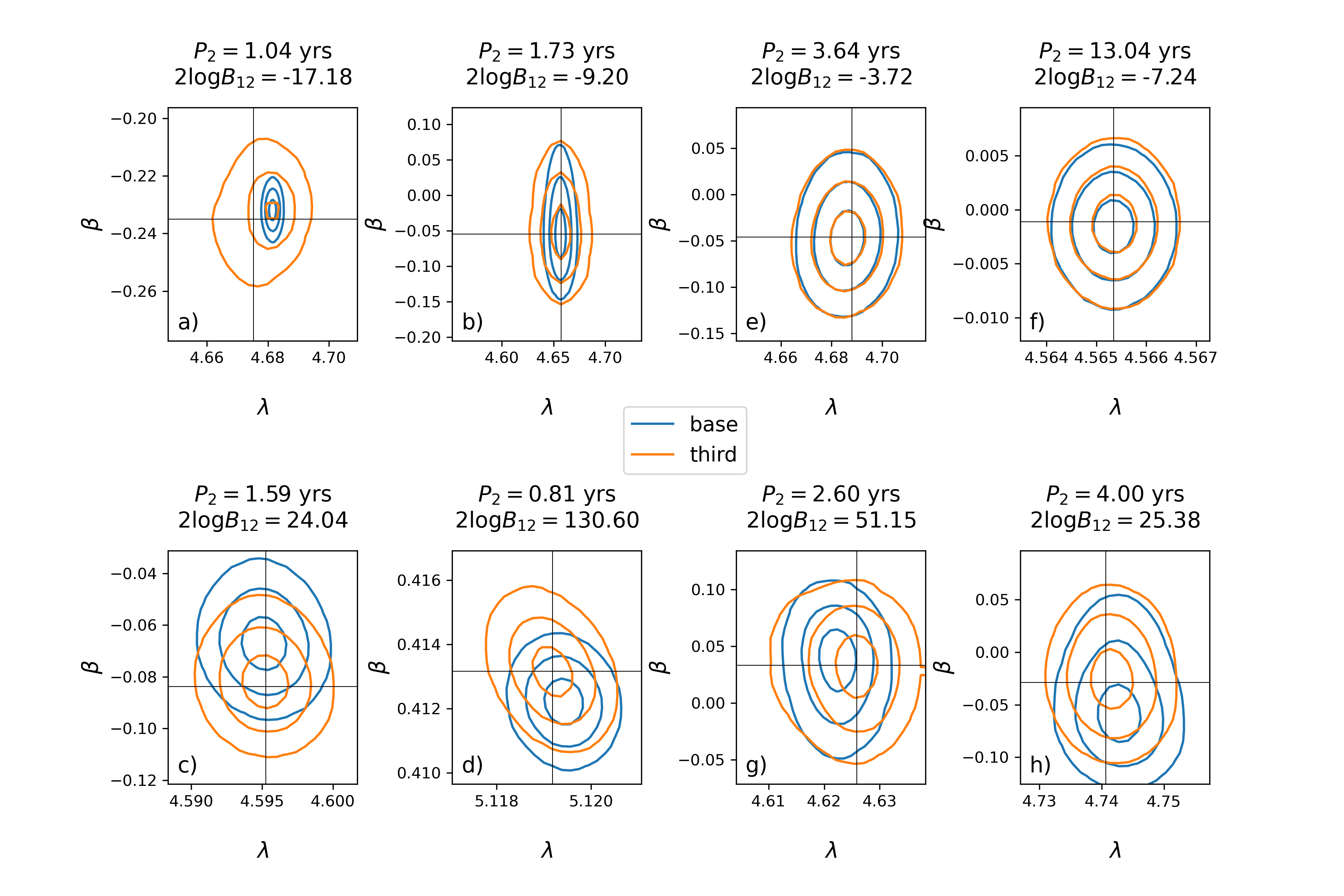}
\caption{Marginalized posterior distributions in the sky localization. The setup is the same as Figure~\ref{fig:f_fdot} with undetectable CBP sources in the top row, detectable sources in the bottom row, short-period ($P_2<T_\text{obs}/2$) systems in the left two columns, and long-period systems ($P_2>T_\text{obs}/2$) in the right two columns. The values of $P_2$ and $2\log{B_{12}}$ are given for each system above each subplot. The posteriors for the base template and third-body template are shown in blue and orange, respectively. The true parameters are shown with the thin black vertical and horizontal lines. All MCMC runs are performed by first injecting the third-body template into the data stream. Undetectable short-period sources center on the mean of the base-template distribution and can be wider than the base-template distribution due to the confusion of the third-body effect with the doppler shift caused by the {\it LISA} motion that aids in probing the ecliptic longitude. The undetectable long-period sources have effectively identical distributions between the two types of templates. Detectable short- and long-period CBP systems usually maintain a similar character between the two different templates, with a bias on the mean of the base-template distributions.}\label{fig:sky_diffs}
\end{center}
\end{figure*}

The sky localizations for the same set of sources in Figure~\ref{fig:f_fdot} are shown in Figure~\ref{fig:sky_diffs}. For the undetectable short-period sources ($a$ and $b$), the third-body template distributions are wider in the ecliptic longitude due to confusion of the doppler shift from {\it LISA}'s motion with the doppler shift from the third-body perturbation. This effect is accompanied by the bias of the third-body distribution towards the center of the base template distribution, as seen in the undetectable $f_0-\dot{f}_0$ posteriors. Detectable short-period sources ($c$ and $d$) have roughly the same size posteriors between the two templates with the base template showing some bias from the true parameters. This is also the case for detectable long-period sources ($g$ and $h$). The long-period systems that are undetectable ($e$ and $f$) have effectively identical sky distributions between the two templates.

\section{Discussion}\label{sec:discussion}

GBs containing a companion perturbing object, such as a BD or CBP, lead to many interesting effects on the posterior distributions of these sources. When examining these effects, three different aspects must be considered: \textit{i}) the effect on individual source posterior distributions; \textit{ii}) the effect on the global fitting of GB sources in terms of their posterior estimation, as well as the overall convergence of the global fitting algorithm; and \textit{iii}) the effects at the population level and the astrophysical implications of these systems. This paper strictly focused on examing (\textit{i}) in detail, but here we build on our findings to consider expectations for all three important aspects.

\subsection{Individual source analysis}

For an individual source, the inclusion of the third-body effects are extreme: the posteriors become much more complicated, especially for sources with long third-body periods. At a basic level, this means the time necessary to converge to a final posterior distribution can be greatly increased. It was routinely seen throughout this work the extreme difference (5-10$\times$) in the time necessary for the two different templates to converge to their final posterior states. Additionally, important quantities such as the frequency, frequency derivative, and the sky localization of these sources can become biased, even when using the correct template if the third-body source is undetectable.

When working to determine the chirp mass and distance to a system, it is required that the frequency derivative be constrained. As previously mentioned, including third-body perturbers tends to select the higher frequency and frequency derivative sources from the catalogue for having a noticeable effect on the waveform phasing. This effect on the phasing means the frequency derivative of these sources is generally constrained. Biases on the frequency and its derivative can lead to a biased determination of the chirp mass and distance to the system. This is important because the `chirping' sources with measureable frequency derivative are considered the most astrophysically interesting: these systems contain more information to help draw conclusions on any underlying astrophysical effects on the inner binary.

A bias on the sky localization may affect electromagnetic measurements of these sources depending on how large the bias is and the specific characteristics of the EM observatory. The biases observed on the sky location of the CBP third-body sources are not large, indicating this should not be a major issue. If the sources are roughly located where predicted, a telescope may locate them. This may help to further constrain the other parameters by refining the priors on its sky localization and any other parameters than can roughly be determined from the electromagnetic measurements, such as the frequency, frequency derivative (if observed long enough), and inclination. 

These individual source aspects can be similarly expected for the much larger number of BD systems. The BD systems contain more mass in the perturber indicating their effective biases are likely to be stronger. However, systems with stronger effects will lead to less confusion over detectability, potentially causing issues in a relatively smaller number of systems compared to the CBP population. This will have to be investigated in more detail in future work.

\subsection{Astrophysical population analysis}

We expect undetectable CBP third-body systems to have little impact on the overall population modelling tests of two-body GB populations because the number of sources is low compared to the two-body GB population and the biases are small relative to the full prior domain over which the population tests will be performed. We also expect this to be the case with any detectable sources as these should be inherently separated out from the population of two-body systems. BD perturbers will likely behave similarly with their stronger effect ensuring the posterior estimates will favor the third-body system.

More work will be needed to understand the modelling of the third-body population and its residual astrophysical uncertainties it leaves behind in the two-body population. Here, we have analyzed an optimistic astrophysical scenario with the underlying astrophysical catalogue allowing us to study the largest diversity of potential sources. More pessimistic catalogues will deal with an overall smaller effect on the underlying population, decreasing the overall impact on parameter estimation and population analysis.

This work has taken an important step toward third-body population analysis with multiple catalogues by building a pipeline for analyzing these large and diverse populations in the most computationally efficient manner. Analyzing the entire population from the beginning with full Evidence estimates would be computationally intractable and wasteful. This pipeline, that proceeds through the stages described in Section~\ref{sec:bayes}, minimizes the necessary computational resources while ensuring that no detectable sources are missed. The undetectable sources do need to be examined though in some detail to confirm their expected minimal effect on the two-body GB population analysis.

Techniques learned over the course of developing, checking, and performing the Evidence computations are also ready to be run on a larger population, which will make this extendable to the full `potentially-detectable' BD third-body catalogue. It can also be used directly for other types of GB astrophysical models to determine the detectability of different effects compared to GR-only driven inspirals.

\subsection{Effects on the global fit}

Global fitting is already challenging even when using basic GB models \citep[e.g.,][]{Littenberg:2020bxy}. In context of the kind of time evolving analysis described in \cite{Littenberg:2020bxy} where the solution is built up as new data arrives, triple systems are likely to be first picked up using the standard two-body templates, with the effects of the third body becoming discernible as more data is accumulated. The inclusion of all of the different potential GB astrophysical effects in the global fit will have to be gathered in steps over time as different effects are suggested and their associated waveforms produced. Each effect will have to be studied at the individual source and population level to further understand the prevalence of each model and its effect on the global fit.

The third-body perturber is clearly a complicated effect and will add confusion and computational burden to the global fitting effort. The overall effect, as well as the effect from the CBP population only, will have to be analyzed with more population models in the future to better understand the global fitting of these systems. With that said, if the actual {\it LISA} data is representative of a more optimistic population, there may be a strong effect on the global fitting procedure. 

As discussed above, the parameters extracted for the undetectable third-body sources may be biased, but it is unlikely that the global fit will choose the third-body template over the base template with regularity while sampling in this case. Similarly, the third-body template is expected to be favored with regularity in the global fit for detectable sources. Systems that are near the detectability threshold may cause more confusion because the algorithm adopted for the global fit will have to operate `switching' between these templates as the sampling proceeds. This will be easier for systems with shorter periods where the $f_0-\dot{f}_0$ remains Gaussian for the third-body template. An improvement would be to define new MCMC proposals that would aid the transition between the two templates, which is usually a nuanced task.

\subsection{A note on an astrophysical $\ddot{f}$ template}

Another template parameterization that includes general astrophysical effects is a template with $\ddot{f}$ as a free parameter rather than the value determined from GR. This template is useful for a subclass of astrophysical effects that chirp smoothly enough to be well-approximated by a quadratic in frequency. Astrophysical tests where this template may be useful include constant acceleration due to a third-body \citep{Tamanini:2019usx,Inayoshi:2017hgw,Bonvin:2016qxr,Wong:2019hsq,Randall:2018lnh}; matter accretion and dynamical friction \citep{Sberna:2020ycl,Kremer:2017xrg,Caputo:2020irr,Cardoso:2019rou}; and tests of gravity \citep{Damour-G, Barausse:2016eii}. In cases where the effect cannot be approximated as a quadratic, the $\ddot{f}$ template will at worst fit as well as the base template due to its one-higher degree of freedom. 

Any (detectable) shorter period third-body source where the orbit is even close to reaching its full angular range will not be well approximated with a quadratic. In our detectable CBP population, for all sources but the longest period source ($P_2\sim17$\,yr), the $\ddot{f}$ template does not perform much differently than the base template. For the longest period source, the $\ddot{f}$ template performs marginally better, but this difference is highly dependent on exactly where in its orbit the third body is when the observation is performed. For BD systems with larger masses, longer periods will be observable with a slower change in the doppler shift. In some initial investigations, we have observed some of these instances showing the $\ddot{f}$ templates fit the third-body effect equally well to the actual third-body template. Further study will be needed to show if the $f_0-\dot{f}_0-\ddot{f}_0$ triplets are consistent with any other astrophysical effects. If they are not consistent with other effects, this template can also act as a means of detectability of the third-body effect; if they are consistent with other effects, it will be harder to be certain about the longer-period BD systems. Analysis with this generalizable template will also be quite nuanced with its Evidence being computed in a dimensionality (9) much more similar to the base template (8) than the third-body template (13). Confusion between the $\ddot{f}$ template and other specific model templates may also lead to more delay in the convergence of the global fit algorithm if they have similar matches against the data.

\section{Conclusion}\label{sec:conclude}

An interesting potential source for {\it LISA} is a Galactic Binary (GB) with a third body in orbit around it. In this work, we produced the first fully Bayesian analysis on the detection of and posterior estimation for a population of circumbinary sub-stellar objects (SSOs). We employed parallel-tempered MCMC techniques to generate posterior distributions and estimate the Bayesian Evidence ratio via Thermodynamic Integration. This analysis also provided the first examination of these sub-stellar mass sources with eccentricity included in their template parameterization.

The posterior behavior of these unique sources is highly dependent on two aspects: the detectability of the source and the period of the third-body orbit. The period values generally bifurcate into two categories above (long period) and below (short period) $P_2=T_\text{obs}/2$, which represents the Nyquist criterion associated with sampling the full orbit of the perturbing object. Detectable sources ($2\log{B_{12}}\gtrsim5$) with short periods remain fairly Gaussian while long period systems have more unpredictable distributions due to the inadequate sampling of the third-body orbit. Undetectable sources have similar behaviors in terms of remaining Gaussian, but the third-body posterior distributions tend to center on the mean of the posterior of the base GR-only template indicating biased parameter measurements whether or not the true template is used. 

These SSO sources came from an optimistic catalogue in order to provide a larger number of potential sources to examine since we were concentrating on the Evidence-based detection and posterior estimation of these systems. The catalogue, while optimistic, was useful because it allowed us to consider a more realistic population that contained a large number of systems generated within a large-dimensional parameter space.
We stress again that absolute numbers of detections reported here should not be taken as representative of the expected performance of {\it LISA}, but at most as an upper optimistic estimate. Another investigation taking into account the variability of the underlying population of CBPs and BDs orbiting {\it LISA} GBs is underway to address astrophysical and population related questions.

The computational cost of this study was considerable. Two routes were taken to ensure tractability. First, the waveform code was reformed for GPU computing capability providing a large acceleration in the necessary Likelihood computations. Second, the process of determining detectability through Evidence ratio estimates and forming full posterior distributions was strategically reserved for only those sources with the potential of third-body detection, as determined through simpler and less time-consuming waveform matching and maximum Likelihood estimation techniques. This allowed us to consider $\sim107$ circumbinary exoplanets systems ($m_c \leq 16\ \Mj$) in detail, 28 of which were detectable. Using these systems as a baseline, we found this optimistic population catalogue also contained a conservative estimate of 954 detectable BDs ($m_c > 16\ \Mj$). 

We note that the qualitative properties of the detected populations found in our study are consistent with those reported in \cite{Tamanini:2018cqb} and \cite{Danielski:2019rvt}; however, a direct comparison of the {\it LISA} detection efficiency was not possible given the new updated features we included in the injected SSO population.

To further analyze the posterior question related to triple GB systems, our study must be expanded with BDs to the posterior level and then in general to other larger objects that may participate in similar triple systems with an inner GB pair. For these larger mass systems, it will be particularly interesting to re-examine the longer period regime. 

Once the initial posterior estimation is understood, studies of these sources across populations will be needed to inform their potential range of prevalence and all of the effects that come with that quality. The first study of this type will be on circumbinary exoplanet systems obtained from different scenarios, as this will be the most computationally reasonable and builds entirely from tools designed here. With both posterior distribution and population studies completed, everything will have to be combined and implemented within the full {\it LISA} global fit framework. 

The overall procedure presented in this work helps provide an initial sense of source detectability, as determined by full Bayesian methods, and potential effects on the overall global fit when including Galactic Binaries with generic astrophysical prescriptions. This process lays the groundwork for expanding to new and different astrophysical models related to these binary sources, providing a road map to ensure the success in the global fitting of generic Galactic Binary systems with {\it LISA}.

\section*{Acknowledgements}
The authors thank Stas Babak and Antoine Petiteau for helpful initial discussions on this project and Silvia Toonen for providing us with a DWD binary population synthesis model. M.L.K. thanks Jonathan Gair and Lorenzo Speri for helpful discussions. C.D. thanks Dimitri Veras for the helpful discussions. 
V.K. acknowledges support from the Netherlands Research Council NWO (Rubicon 019.183EN.015 grant).
N.K. acknowledges the support from the Gr-PRODEX 2019 funding program (PEA 4000132310).
This research was supported in part through the computational resources and staff contributions provided for the Quest/Grail high performance computing facility at Northwestern University.
N.T.~acknowledges support form the French space agency CNES in the framework of {\it LISA}. C.D. acknowledges financial support from the State Agency for Research of the Spanish MCIU through the `Center of Excellence Severo Ochoa' award to the Instituto de Astrof\'isica de Andaluc\'ia (SEV-2017-0709), and the Group project Ref. PID2019-110689RB-I00/AEI/10.13039/501100011033. This work was supported by the COST action CA16104 ``Gravitational waves, black holes and fundamental physics'' (GWverse) through a Short Term Scientific Mission (STSM) awarded to C.D. (ID 48268) and V.K. (ID 48272). COST actions are supported by COST (European Cooperation in Science and Technology). N.J.C. appreciates the support of the NASA LISA Foundation Science grant 80NSSC19K0320.
This project has received financial support from the CNRS through the MITI interdisciplinary programs.
This paper also employed use of {\sc SciPy} \citep{scipy} and {\sc Matplotlib} \citep{Matplotlib}.

\section*{Data Availability}

The waveform codes used in this work are from the publicly available {\sc gbgpu} package \cite{michael_l_katz_2022_6500434}. This package is based on the original fast GB code from \citet{Cornish:2007if}  and the third-body-inclusive fast GB code used in \citet{Robson:2018svj}. The maintained and documented version of the new code can be found on Github (\url{https://github.com/mikekatz04/GBGPU}). The specific code version used in this paper is from an older iteration of {\sc gbgpu} that can be accessed through the above Github repository  \href{https://github.com/mikekatz04/GBGPU/tree/b65f281c501509c6ad25118d62d447eb009593d5}{here}. The code makes use of {\sc CuPy}~\citep{CuPy}, {\sc NumPy}~\citep{Numpy}, {\sc Cython}~\citep{Cython}, and a special {\sc Cython} wrapper for CUDA from \citet{CUDAwrapper} to parallelize the waveform and Likelihood computations for GPUs and then wrap them into {\sc Python} so they are accessible to the other MCMC and analysis codes used here.  The parallelization model involves batching many waveform and Likelihood computations together at fixed wall time, therefore, decreasing the per-Likelihood cost substantially. During the MCMC runs, typically thousands of Likelihoods are computed simultaneously reducing the per-Likelihood computational cost to $\lesssim5\mu$s on an NVIDIA A100 GPU. Please note the codes are written for use on both CPUs and GPUs. For reference, the general per-Likelihood computation time on a CPU is $\sim1$ms.


The code for our MCMC sampler ({\sc Eryn},~\citeauthor{karnesis-in-prep}) is available upon request to the authors and will eventually be made public. 

The search, parameter estimation, and Evidence computations required the storage of a large amount of data. This data is also available by request to the authors.



\bibliographystyle{mnras}






\bsp	
\label{lastpage}
\end{document}